\documentclass[aps,prb,twocolumn,showpacs,amsmath,amssymb,superscriptaddress]{revtex4-2}
\usepackage{graphicx}
\usepackage{dcolumn}
\usepackage{bm}
\usepackage{color}
\usepackage{hyperref}
\usepackage{float}
\hypersetup{pdfnewwindow=true, colorlinks=true, linkcolor=blue, anchorcolor=blue, citecolor=blue, filecolor=blue, menucolor=blue, urlcolor=blue}

\begin{document}

\title{Grotthuss-type oxygen hole polaron transport in desodiated Na$_{2}$Mn$_3$O$_7$}

\author{Ming Lei}
\email{mlei012@mit.edu}
\affiliation{Department of Materials Science and Engineering, Massachusetts Institute of Technology, Cambridge, MA 02139, USA}
\author{Iwnetim I. Abate}
\email{iabate@mit.edu}
\affiliation{Department of Materials Science and Engineering, Massachusetts Institute of Technology, Cambridge, MA 02139, USA}

\begin{abstract}

Polarons are quasiparticles that arise from the coupling of electrons or holes with ionic vibrations in polarizable materials. Typically, they are either localized at a single atomic site or delocalized over multiple sites. However, after the desodiation of Na$_{2}$Mn$_{3}$O$_{7}$, we identify a rare split-hole polaron, where a single hole is shared between two adjacent oxygen atoms rather than fully localized or delocalized. We present a density functional theory (DFT) study on the migration and transport properties of these oxygen hole polarons in NaMn$_{3}$O$_{7}$ and Na$_{1.5}$Mn$_{3}$O$_{7}$. Our calculations reveal that the split polaron configuration near a sodium vacancy is the ground state, while the localized polaron acts as the transition state. Migration occurs via a stepwise charge transfer mechanism along the $b$-axis, where the split-hole polaron transitions through a localized hole state. This transport behavior closely resembles the Grotthuss mechanism, which describes proton transport in H$_{2}$O. We compute the polaron mobility as $\mu$ = 1.37 $\times$ 10$^{-5}$ \text{cm}$^2$/\text{(V$\cdot$s)} with an energy barrier of 242 \text{m\text{eV}}. Using the Mulliken-Hush theory, we determine the electronic coupling parameter $V_{AB}$ = 0.87 \text{eV}. A similar migration mechanism is observed in Na$_{1.5}$Mn$_{3}$O$_{7}$, where the split polaron remains more stable than in the localized state. This study provides the first theoretical investigation of split-hole polaron migration, offering new insights into the charge transport of exotic polaronic species in materials with implications for a wide range of functional materials including battery cathodes, thermoelectrics, photocatalysts, and next-generation optoelectronic devices.
\end{abstract}

\maketitle

\section{Introduction}
\begin{figure*}[ht]
\centerline{\includegraphics[width=\textwidth]{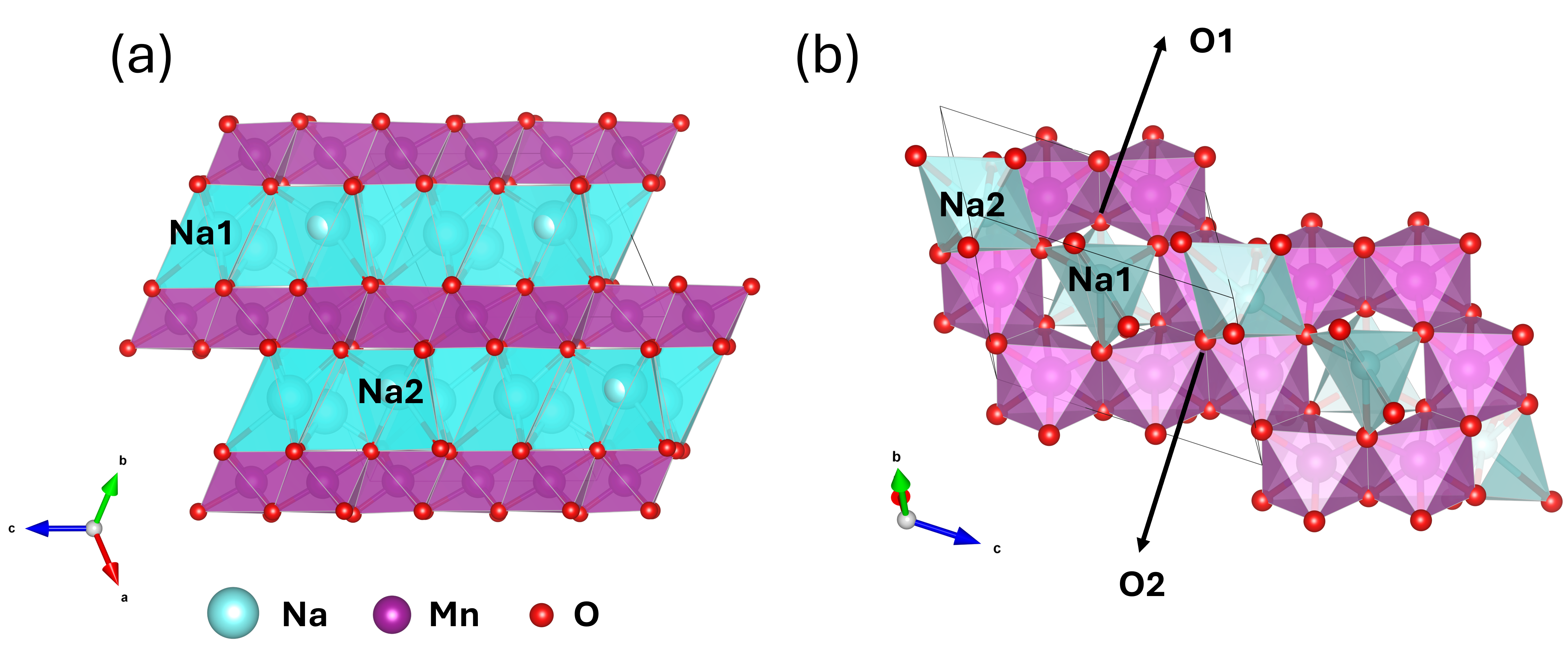}}
\caption{\label{fig:n2mo}(a) Side view and (b) top view of the Mn--O layer structure in Na$_{2}$Mn$_{3}$O$_{7}$. Mn atoms are shown in magenta, O atoms in red, and Na atoms in blue. Na atoms located beneath the Mn vacancy layer are labeled Na1; they occupy trigonal prismatic sites in the layers above and below the Mn vacancies. In contrast, Na atoms situated farther from the vacancy layer are labeled Na2 and are octahedrally coordinated. Oxygen anions bonded to two Mn atoms are labeled as O1, and those bonded to three Mn atoms are labeled as O2.}
\end{figure*}

In many polarizable materials, the motion of electrons or holes is accompanied by local lattice distortions, forming polarons. In oxide materials, polarons often appear as narrow-localized electronic states in the middle gap\cite{franchini,Schirmer} In real space, polarons can be highly localized at a single atomic site or delocalized over multiple sites. For example, in an octahedral symmetry, an oxygen hole polaron may be confined to one site or spread across six oxygen atoms. However, in Na$_{2}$Mn$_{3}$O$_{7}$ (a material with a rare ordered-vacancy motif, also known as the maple leaf lattice, Fig.~\ref{fig:n2mo}), Abate et al.\cite{D1EE01037A} discovered an unusual polaron configuration in which the hole is neither fully localized nor completely delocalized upon desodiation. Instead, a single hole is shared between two adjacent oxygen atoms, a phenomenon that we term a split-hole polaron, indicating the division of a hole into two partial charges. In this desodiated state, two sodium vacancies per unit cell can accommodate two oxygen hole polarons. Density functional theory (DFT) calculations reveal that the split-hole polaron configuration is energetically more stable than the fully localized state by 242 \text{m\text{eV}}. Although this is the first reported case of its kind (particularly as it arises from desodiation and electrochemically), similar polaron states have been identified in BaTiO$_{3}$ through Electron Paramagnetic Resonance spectroscopy, where a hole is shared between two neighboring oxygen atoms near substitutional defects, such as Al substituting for Ti (Al$_\text{Ti}$).\cite{Possenriede} Computational studies further confirm that such hole states can be stabilized through electron correlations and defect-induced lattice deformations.\cite{Donnerberg}

Numerous computational studies have extensively explored the migration mechanisms of localized electron and hole polarons, providing valuable insights into their transport dynamics. Rosso\cite{Rosso} investigated the migration of electrons in hematite ($\alpha$-Fe$_2$O$_3$) by calculating the migration barrier along a pathway generated by linear interpolation between the initial and final geometries. This linear interpolation approach has also been widely applied to study electron and hole polaron migration in other materials, including olivine Li$_x$FePO$_4$,\cite{Maxisch} nontronite Fe$_2$Si$_4$O$_{10}$(OH)$_2$,\cite{Alexandrov} perovskite SrTiO$_3$,\cite{Naoto} anatase and rutile TiO$_2$,\cite{Dupuis,Palermo} and BiVO$_4$.\cite{Michel} In this work, we employ DFT to investigate the migration mechanism of the split-hole polaron in NaMn$_3$O$_7$. We calculate the electronic coupling matrix element $V_{AB}$ and the mobility of the oxygen hole polaron transport. Furthermore, we demonstrate that a similar split-hole polaron can also be observed in the less desodiated phase, Na$_{1.5}$Mn$_3$O$_7$, exhibiting transport behavior consistent with that observed in NaMn$_3$O$_7$.

\section{Calculation Method}
Spin-polarized structural optimization and total energy calculations were performed using the Vienna Ab initio Simulation Package (VASP 5.4.4).\cite{Kresse} Since the Generalized Gradient Approximation (GGA) of Perdew, Burke, and Ernzerhof (PBE)\cite{Perdew}, combined with the Hubbard U correction (PBE+U),\cite{plusU} fails to accurately predict the relative energies of localized and split oxygen hole polarons, also PBE+U alone either converges to a fully delocalized solution or spontaneously breaks symmetry to yield a localized state,\cite{D1EE01037A} the exchange-correlation (XC) effects were instead treated using the HSE06 screened hybrid functional.\cite{Heyd} A plane-wave energy cutoff of 500 \text{eV} was employed, along with projector-augmented wave (PAW) potentials using the following valence electron configurations: Na: [Ne] 3s$^1$, Mn: [Ar] 4s$^1$3d$^6$, and O: [He] 2s$^2$2p$^4$. The integration of the Brillouin zone was performed using a 4$\times$4$\times$4 $\Gamma$-centered k-point mesh. Phonon calculations were performed on a 2$\times$2$\times$2 supercell, and the structure of the phonon band was analyzed using the PHONOPY package.\cite{TOGO20151}

\section{results and discussions}

\subsection{Oxygen hole polaron migration mechanism in NaMn$_3$O$_7$}

Electrochemical desodiation of Na$_{2}$Mn$_{3}$O$_{7}$ removes both a Na$^+$ ion and an electron, forming hole-doped NaMn$_{3}$O$_{7}$. Owing to the ordered-vacancy nature of the structural motif, Na$_{2}$Mn$_{3}$O$_{7}$ exhibits two crystallographically distinct local environments for both Na and O atoms (Fig.~\ref{fig:n2mo}). Sodium atoms residing beneath the Mn vacancy layer are designated as Na1 (occupy trigonal prismatic sites in the layers above and below the Mn vacancies), while those situated away from the vacancy layer are referred to as Na2 (octahedrally coordinated). First-principles calculations indicate that Na1 sites are thermodynamically more stable than Na2; thus, electrochemical desodiation proceeds via preferential extraction of Na$^+$ ions from Na2 sites.\cite{C7TA02609A} Analogously, oxygen atoms adjacent to the Mn vacancy sites (denoted O1) are undercoordinated, possessing non-bonding lone pairs, in contrast to fully coordinated oxygen atoms located farther from the vacancies (O2). As a consequence, hole doping upon desodiation is energetically favored on the O1 sublattice. The holes do not localize on Mn sites due to their negligible contribution to the electronic density of states near the Fermi level, which is predominantly composed of O 2$p$ orbitals associated with O1 atoms (a characteristic feature of negative charge-transfer materials).\cite{D1EE01037A}

\begin{figure*}[ht]
\centerline{\includegraphics[width=\textwidth]{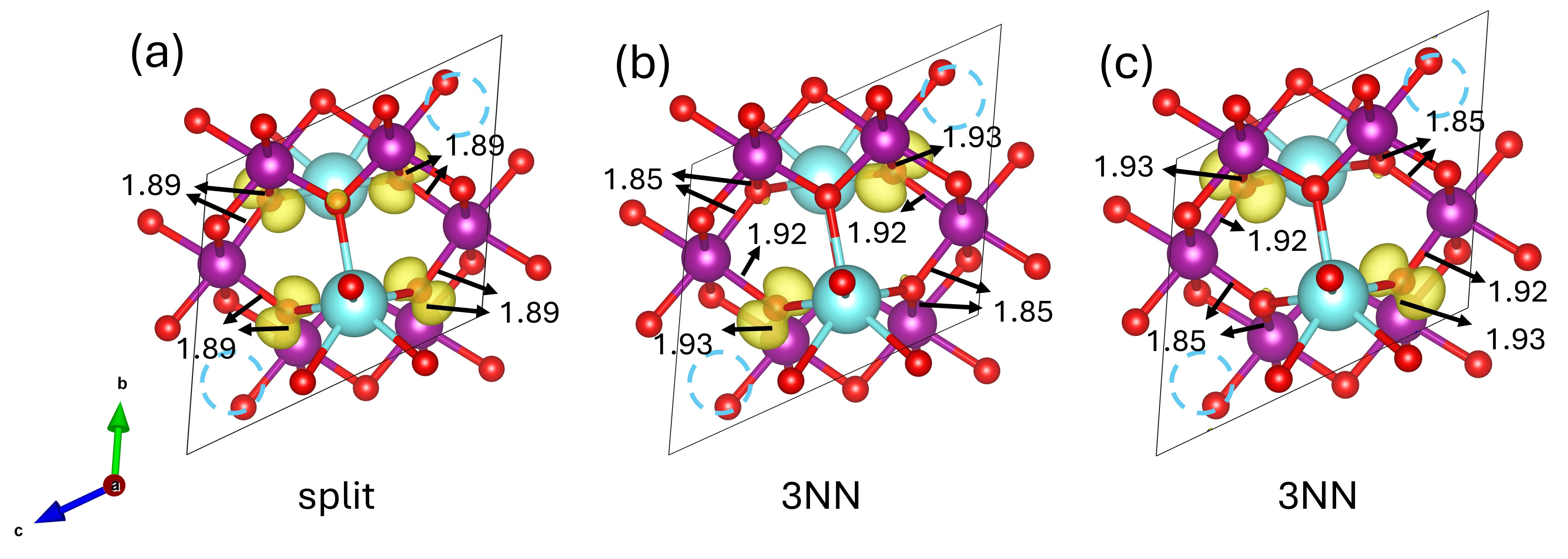}}
\caption{\label{fig:3nn}(a) Split oxygen hole polaron configuration. (b) and (c) two energetically equivalent 3NN localized oxygen hole polaron configurations in NaMn$_3$O$_7$. The two 3NN configurations differ in the direction of hole localization relative to the Mn vacancy: in (b), the hole resides on the lower-left and upper-right sites within the local coordination environment, while in (c), it is on the upper-left and lower-right sites. The Mn--O bond lengths (in \AA) are indicated, and spin density is shown in yellow dumbbells. Mn atoms are shown in magenta, O atoms in red, and Na atoms in blue. The Na vacancies are shown in dashed blue circles.}
\end{figure*}

Abate et al.\cite{D1EE01037A} demonstrated hole doping of O1 atoms upon desodiation through a combination of DFT and advanced spectroscopic techniques, including resonant inelastic X-ray scattering and X-ray absorption spectroscopy. To identify the most stable hole configuration, various structural models were evaluated, among which the split-hole polaron and the localized hole emerged as key candidates (Fig.~\ref{fig:3nn}). In the localized oxygen hole polaron configuration, when the two holes are maximally separated within the manganese vacancy center (forming a linear chain consisting of an oxygen polaron, a Mn vacancy, and another oxygen polaron) it is referred to as the three nearest-neighbor (3NN) configuration. There are two energetically equivalent 3NN states, illustrated in Figs.~\ref{fig:3nn}b and c. In one configuration, the oxygen hole polarons are located on the top-right and bottom-left oxygen atoms. For these oxygen atoms, the Mn--O bond lengths are elongated, ranging from 1.92 to 1.93 \AA. In contrast, for the two oxygen atoms that do not host hole polarons, the Mn--O bond lengths are shorter, around 1.85 \AA. Due to structural symmetry, a second equivalent 3NN configuration exists in which the hole polarons are located on the top-left and bottom-right oxygen atoms, exhibiting the same Mn--O bond length pattern.

To investigate the migration pathway of the oxygen hole polaron, we performed a linear interpolation\cite{Rosso,Maxisch, Alexandrov, Naoto, Dupuis, Palermo, Michel} between two symmetry-equivalent 3NN localized configurations, where the hole resides on either the lower-left and upper-right or the upper-left and lower-right sites within the local coordination environment around the Mn vacancy. The midpoint of this linear interpolation is the split-hole configuration (Fig.~\ref{fig:3nn}a), where the hole is delocalized over four oxygen atoms. This configuration is energetically favored, lying 242~meV lower in energy than the 3NN localized states. These results indicate that the split-hole configuration is the thermodynamic ground state in NaMn$_3$O$_7$, while the 3NN localized configurations act as transition states. The corresponding energy profile is shown in Fig.~\ref{fig:barrier}, and structural details of the intermediate images are provided in the Supplemental Material.\cite{supp} We note that the intermediate structures along the interpolation path were not locally relaxed. The fully relaxation of these configurations always relax to either the 3NN or split states, indicating the absence of other metastable minima. Although local relaxation could slightly modify the shape of the energy profile, it would not affect the location or energy of the transition state. The linear interpolation approach thus provides a physically meaningful and computationally efficient estimate of the migration barrier.

This naturally raises the question: How does the split polaron migrate in NaMn$_3$O$_7$? In the following, we examine the oxygen hole polaron migration mechanism of the split polaron in detail. 

\begin{figure}
\centerline{\includegraphics[width=3.5in]{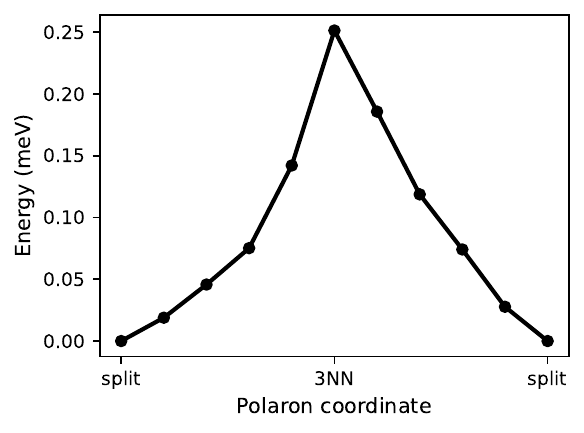}}
\caption{\label{fig:barrier}Energy profiles for oxygen hole polaron transport in NaMn$_3$O$_7$.}
\end{figure}

\begin{figure*}[ht]
\centerline{\includegraphics[width=\textwidth]{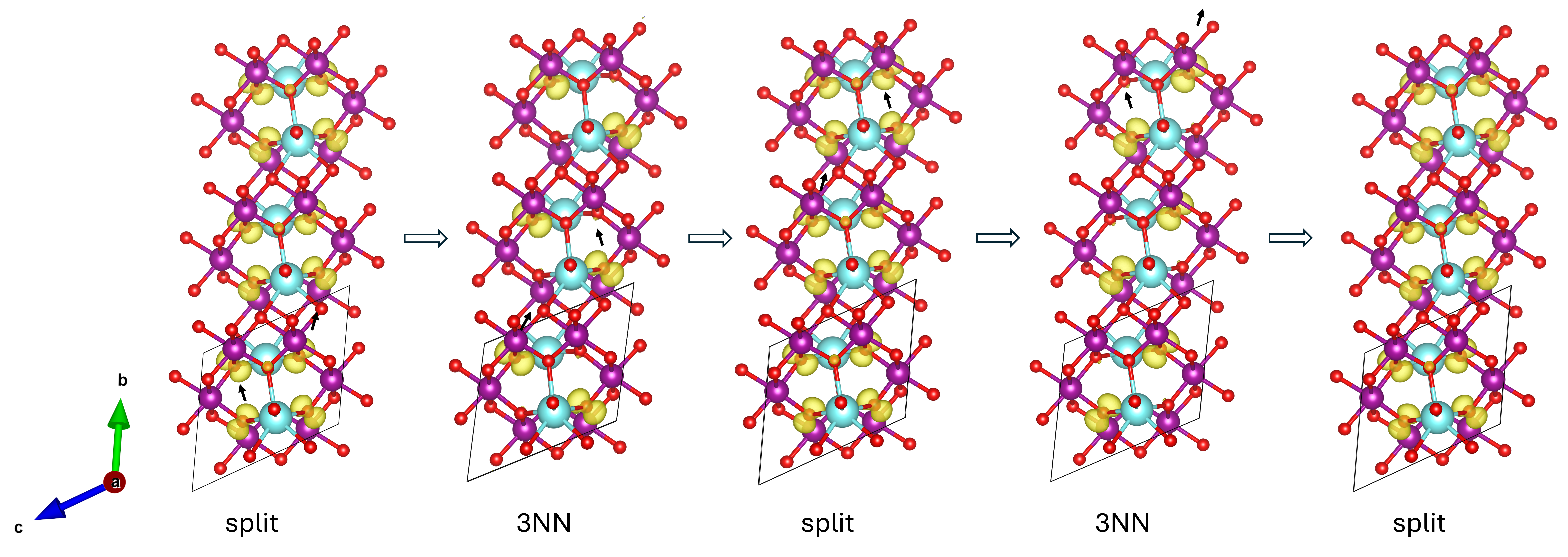}}
\caption{\label{fig:hopping}The migration mechanism of the oxygen hole polaron in NaMn$_3$O$_7$. Those arrows denote the migration direction of oxygen hole polarons. The sequence shown represents one representative migration path. The diagonal orientation of each 3NN transition state may vary randomly among symmetry-equivalent options without affecting the overall transport behavior.}
\end{figure*}

The migration of oxygen hole polarons in NaMn$_3$O$_7$ can be described as a stepwise hopping process along the crystallographic $b$-axis, as illustrated in Fig.~\ref{fig:hopping}. This migration involves the sequential redistribution of a polaronic charge among neighboring oxygen atoms, facilitated by the structural arrangement of Na vacancies and oxygen coordination. In principle, the oxygen hole can migrate through either of the two symmetry-equivalent 3NN configurations in each step. The observed alternation in diagonal orientation of each 3NN
transition state is therefore arbitrary and it depends on which of the two equivalent paths is selected at each hop.

\begin{figure}[h]
\centerline{\includegraphics[width=3.5in]{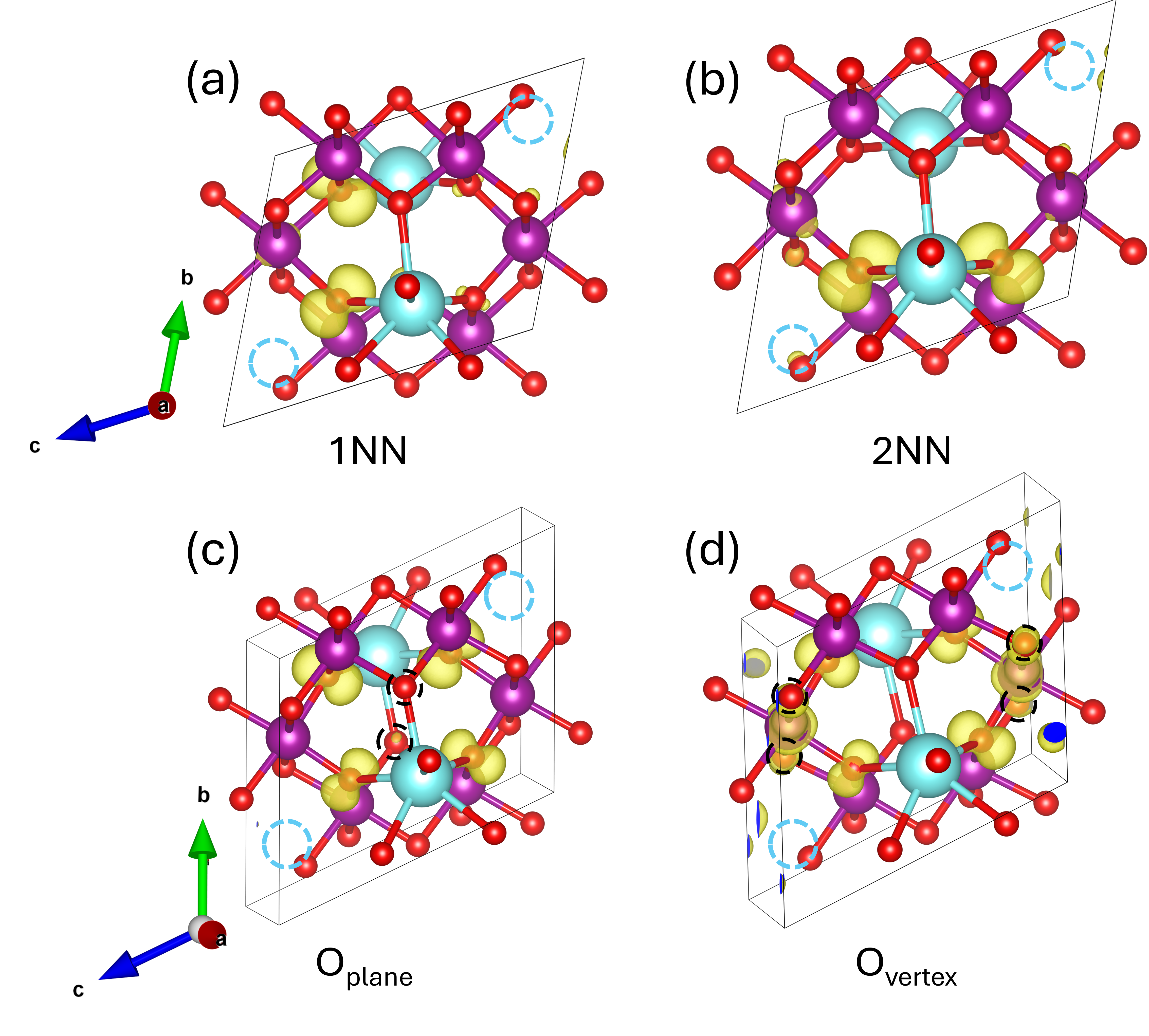}}
\caption{\label{fig:plane}Other possible oxygen hole polaron configurations. (a) 1NN configuration: two oxygen hole polarons located on first-nearest-neighbor O atoms. (b) 2NN configuration: two oxygen hole polarons located on second-nearest-neighbor O atoms. (c) A hole polaron localized on the O$_\text{plane}$ atoms. (d) A hole polaron localized on the O$_\text{vertex}$ atoms. The black dashed circles in (c) and (d) indicate the O$_\text{plane}$ and O$_\text{vertex}$ atoms, respectively.}
\end{figure}

Initially, the polaron exists in a split configuration, where the hole polarons are delocalized between two oxygen atoms near a sodium vacancy. During the migration, the polaron progressively shifts from the initial oxygen pair to the adjacent oxygen atom, passing through an intermediate state where the polaron is more localized on two oxygen atoms. This stepwise hole polarons redistribution preserves the structural symmetry and minimizes disruption to the Mn–O bonding network. Such preservation of the bonding network lowers both elastic strain and electronic reorganization energy during migration. Since the polaron resides primarily on the oxygen sublattice and hybridizes with Mn $d$ orbitals, maintaining the integrity of the Mn–O framework helps stabilize the charge distribution and contributes to the low-energy hopping pathway.

This stepwise polaron transfer process is similar to the Grotthuss mechanism\cite{de1806decomposition} for proton transport in water, where a proton propagates through a hydrogen-bonded network by transiently delocalizing its charge across multiple water molecules. Likewise, in NaMn$_3$O$_7$, the polaron migrates by sequential hole transfer between oxygen atoms, involving temporary hole localization during the migration process. However, a key distinction lies in the medium: while the Grotthuss mechanism occurs in a dynamic liquid environment with flexible hydrogen bonds, polaron migration in NaMn$_3$O$_7$ takes place in a rigid crystalline lattice. The structural symmetry and fixed positions of the Na vacancies guide the migration pathway, ensuring consistent polaron hopping along the $b$-axis with minimal structural distortion. Therefore, the stepwise redistribution of the oxygen hole polaron, structural constraints, and the energy landscape collectively define the migration mechanism of the oxygen hole polaron in NaMn$_3$O$_7$.

\subsection{\texorpdfstring{Other possible oxygen hole polaron migration transition states in NaMn\texorpdfstring{$_3$}{3}O\texorpdfstring{$_7$}{7}}{Other possible oxygen hole polaron migration transition states in NaMn3O7}}

As discussed earlier, the localized 3NN polaron configuration serves as the transition state during split polaron migration in NaMn$_3$O$_7$. This raises the question of whether alternative oxygen hole polaron configurations could also act as possible transition states for the migration process.

To investigate this possibility, we first considered the 1NN and 2NN configurations, where the two hole polarons occupy the first-nearest and second-nearest neighbor oxygen atoms, respectively, as shown in Fig.~\ref{fig:plane}a and b. However, relative to the 3NN localized polaron configuration, the 1NN and 2NN configurations are higher in energy by 101meV and 58meV, respectively.\cite{D1EE01037A} These significant energy penalties suggest that neither the 1NN nor 2NN configurations can act as feasible transition states for oxygen hole polaron migration.

Next, we examined a configuration where the hole polaron is localized on two oxygen atoms bonded to two Mn atoms and one Na atom, nearly coplanar. This structure, referred to as O$_{\text{plane}}$ (marked in black dashed circles in Fig.~\ref{fig:plane}c), initially exhibits an Mn--O bond length of 1.84~\AA{} with its neighboring Mn atoms. To encourage hole localization on the O$_{\text{plane}}$ atoms, we introduced structural distortions starting from the split polaron configuration, manually elongating the Mn--O bond length from 1.84~\AA{} to 2.00~\AA{}. However, after full structural relaxation, the structure reverted to the more stable split polaron configuration, indicating the instability of the O$_{\text{plane}}$ state. A static self-consistent field (SCF) calculation (at full elongation of 2.00~\AA{} without structural relaxation) further confirmed that the O$_{\text{plane}}$ configuration is 919~\text{m\text{eV}} less stable than the split polaron state. Moreover, the degree of hole localization on the O$_{\text{plane}}$ atoms was minimal, as shown in Fig.~\ref{fig:plane}c, indicating weak polaronic character and ruling out this configuration as a potential transition state.

Finally, we explored another possible configuration where the hole polaron is distributed across four oxygen atoms with Mn--O bonds oriented perpendicular to the plane of the split polaron, referred to as O$_{\text{vertex}}$ (marked in black dashed circles in Fig.~\ref{fig:plane}d). The initial Mn--O bond length between the O$_{\text{vertex}}$ atoms and their neighboring Mn atoms was 1.91~\AA{}. Similar to the previous test, structural distortions were introduced by stretching the Mn--O bond length from 1.91~\AA{} to 2.10~\AA{} in an attempt to localize the hole polaron on the O$_{\text{vertex}}$ atoms. However, after structural relaxation, the system again reverted to the split polaron configuration. A static SCF calculation revealed that the O$_{\text{vertex}}$ configuration was 1122~\text{m\text{eV}} less stable than the split polaron configuration.

Although hole localization was observed during this distortion, as depicted in Fig.~\ref{fig:plane}d, the large energy penalty rules out this configuration as a viable transition state. To further test the robustness of this conclusion, a smaller distortion was applied, stretching the Mn--O bond length from original 1.91 \AA{} to 2.00 \AA{}. In this case, no hole localization occurred, further supporting that the O$_{\text{vertex}}$ configuration cannot serve as a transition state for polaron migration in NaMn$_3$O$_7$.

In this work, we systematically examined a range of symmetry-distinct polaron configurations—including 1NN, 2NN, O$_\text{plane}$, and O$_\text{vertex}$ states—as potential transition states. Among these, only the 3NN localized configuration exhibited sufficient hole localization and an energy barrier consistent with a transition state. All configurations were selected based on their proximity to the Mn vacancy, where the oxygen hole is most likely to localize. While additional possible transition states may exist, the explored set covers the most physically plausible candidates based on local structural motifs, vacancy-centered hopping mechanisms, and the symmetry of the host lattice. Based on these findings, the 3NN localized polaron configuration remains the most likely transition state, as all other tested configurations show either significantly higher energies or insufficient hole localization to facilitate polaron migration.

\subsection{Oxygen hole polaron transport mobility in NaMn$_3$O$_7$}

\begin{figure*}[ht]
\begin{minipage}{\columnwidth}
\includegraphics[width=3.0in]{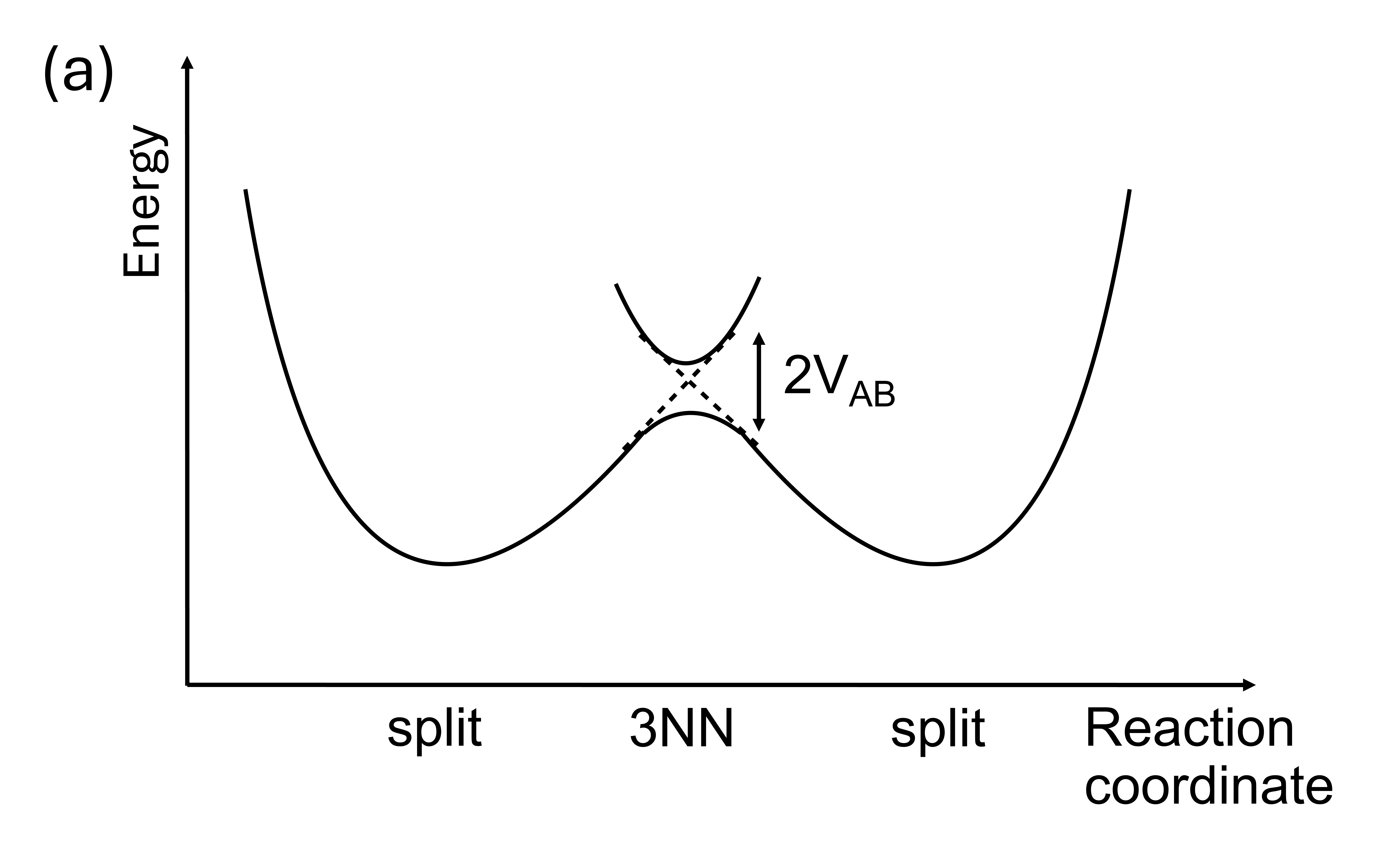}
\end{minipage}
\begin{minipage}{\columnwidth}
\includegraphics[width=3.0in]{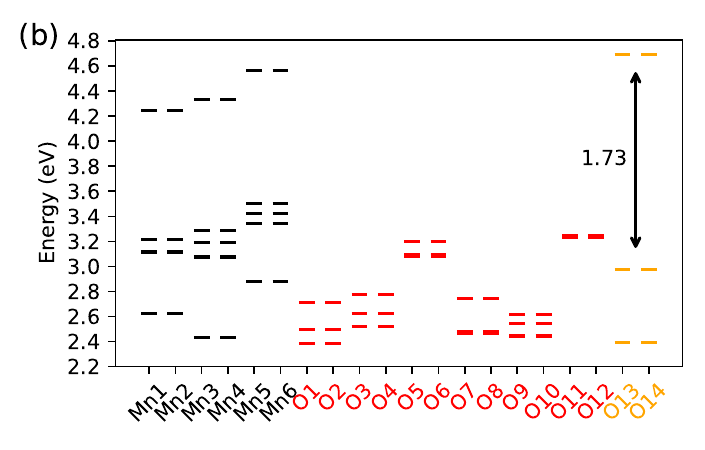}
\end{minipage}
\caption{(a) Schematic showing a cross section of an energy profile for oxygen hole polaron hopping between initial (split) and final (split) states in a typical symmetric hole-transfer reaction. (b) The calculated onsite energies of Mn 3$d$ and O 2$p$ orbitals in the transition state (localized 3NN polaron configuration).}
\label{fig:dos}
\end{figure*}

Having described the oxygen hole polaron migration mechanism, we now focus on the transport mobility of oxygen hole polarons in NaMn$_3$O$_7$. Within Marcus theory,\cite{MARCUS} the potential energy surfaces governing electron or hole transfer are often modeled as quadratic functions of a collective nuclear coordinate q, as illustrated in Fig.~\ref{fig:dos}a. These transfer processes can be categorized as either electronically adiabatic or nonadiabatic, with the rate expressions in both cases primarily determined by two critical factors: the reorganization energy and the electronic coupling matrix element $V_{AB}$.

There are two primary methods to calculate the electronic coupling matrix element $V_{AB}$. The first approach involves molecular cluster calculations using the electron transfer (ET) module in NWChem.\cite{nwchem} This method requires constructing molecular clusters representing both the initial and final states of the polaron transfer. However, since both the initial and final states in NaMn$_3$O$_7$ correspond to the split polaron configuration, this approach is not applicable for NaMn$_3$O$_7$.

The second method, based on Mulliken-Hush theory,\cite{Hush, Adelstein} relates the electronic coupling parameter $V_{AB}$ to the energy difference between the adiabatic bonding and antibonding electronic states at the transition state, denoted as $\Delta E_{12}$, and is expressed as:

\begin{equation}
V_{AB} = \frac{1}{2} \Delta E_{12} 
\end{equation}

To estimate this, we used Wannier functions to calculate the onsite energies of Mn 3$d$ and O 2$p$ orbitals in the transition state, represented by the localized 3NN polaron configuration, as shown in Fig.~\ref{fig:dos}b.  O13 and O14 are the two oxygen atoms that host the holes. The energy difference between the two O 2$p$ orbitals is 1.73~\text{eV}, yielding an electronic coupling matrix element $V_{AB}$ of 0.87 \text{eV}.  

The computed $V_{AB}$ in NaMn$_{3}$O$_{7}$ is large compared to values reported for other small polaron systems. For example, in hematite ($\alpha$-Fe$_{2}$O$_{3}$), the DFT+U studies have reported values ranging from 0.2 to 0.23 \text{eV}.\cite{Iordanova,Adelstein} In TiO$_{2}$, $V_{AB}$ varies from ~0.02~\text{eV} to 0.20~\text{eV} depending on the crystallographic direction,\cite{Dupuis} while in nontronite Fe$_2$Si$_4$O$_{10}$(OH)$_2$, typical values are below 0.06~\text{eV}.\cite{Alexandrov} The significantly larger coupling in our system indicates that the polaron transfer occurs in the adiabatic regime, where charge carriers transit smoothly along a localized 3NN transition state without requiring quantum tunneling. 

Next, we calculated the oxygen hole polaron transport mobility in NaMn$_3$O$_7$. The transport mobility $\mu$ of the oxygen hole polaron can be determined using the Einstein relation:

\begin{equation}
\mu = \frac{eD}{k_BT} 
\end{equation}

where $e$ is the elementary charge, $k_B$ is the Boltzmann constant, $T$ is the temperature (298~K), and $D$ is the diffusion coefficient, given by

\begin{equation}
D = \frac{a^2\tau}{2}
\end{equation}

where $a$ represents the hole polaron transfer distance, measured as 2.665~\AA{} along the $b$-axis, and $\tau$ is the adiabatic transition rate, expressed as:

\begin{equation}
\tau = A \exp\left( \frac{-\Delta E^{\ddagger}}{k_BT} \right)
\end{equation}

Here, $A$ denotes the pre-exponential factor, and $\Delta E^{\ddagger}$ is the activation energy barrier for the polaron transfer (242~meV). The pre-exponential factor $A$ can be determined using:

\begin{equation}
A = n\tau_0
\end{equation}

where $n$ is the number of equivalent neighboring sites to which the hole polaron can hop, set to 2 in this study (there are two energetically equivalent 3NN transition states), and $\tau_0$ is the attempt frequency, calculated following the method proposed by Koettgen.\cite{C6CP04802A}

\begin{align}
\tau_0 &= \dfrac{k_B T}{h} \dfrac{ 
\prod_{q}^{M}\prod_{i}^{N}
\left(2\sinh\left(\dfrac{h\nu_{q,i}}{2k_B T}\right)\right)^{\frac{1}{M}}}
{\prod_{m}^{M}\prod_{j}^{N-1}
\left(2\sinh\left(\dfrac{h\nu_{m,j}}{2k_B T}\right)\right)^{\frac{1}{M}}} \\
&\approx \dfrac{k_B T}{h}\dfrac{ 
\prod_{q}^{M}\prod_{i}^{N}
\left(1 - e^{-\dfrac{h\nu_{q,i}}{k_B T}}\right)^{\frac{1}{M}}}
{\prod_{m}^{M}\prod_{j}^{N-1}
\left(1 - e^{-\dfrac{h\nu_{m,j}}{k_B T}}\right)^{\frac{1}{M}}}
\label{eq:tau}
\end{align}

The indices $i$ (phonon band) and $q$ (wave vector) for the initial state, represented by the split polaron configuration, correspond to the indices $j$ (phonon band) and $m$ (wave vector) for the transition state, described by the localized 3NN polaron configuration. Notably, the transition state has one degree of freedom fewer than the initial state due to the constraint imposed by the polaron localization.		

\begin{figure*}[t]
\begin{minipage}{\columnwidth}
\includegraphics[width=3.5in]{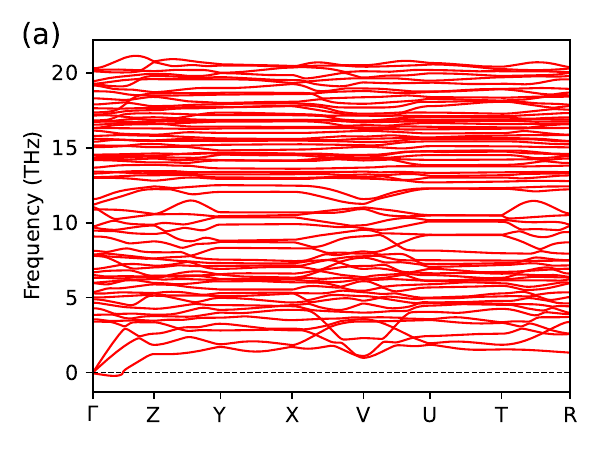}
\end{minipage}
\begin{minipage}{\columnwidth}
\includegraphics[width=3.5in]{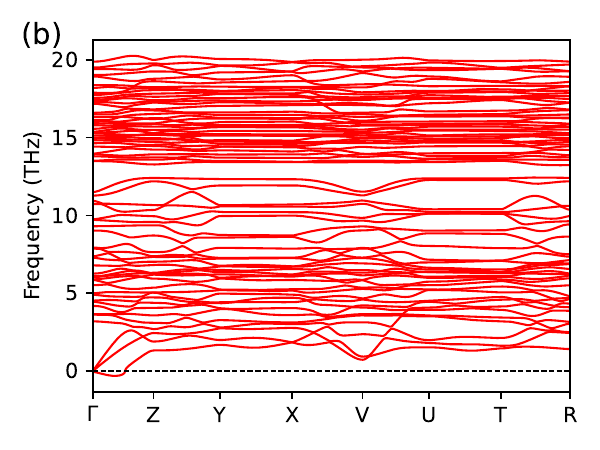}
\end{minipage}
\caption{Phonon dispersions of the (a) transition state (localized 3NN polaron configuration) and (b) initial state (split polaron configuration).}
\label{fig:phonon}
\end{figure*}

The calculated phonon dispersions for both the transition and initial states are presented in Fig.~\ref{fig:phonon}. There is a soft acoustic mode near the $\Gamma$ point with a small negative frequency. The largest imaginary frequency near the $\Gamma$ point is $-0.24$~THz, which is quite small. Given that we are using a $2\times2\times2$ supercell, this is likely a finite-size effect rather than a real instability. Increasing the supercell size might further reduce this, but the computational cost would be significantly higher. Additionally, from Eq.~\ref{eq:tau}, the attempt frequency ($\tau_0$) is calculated as a product of phonon mode contributions. Since this small imaginary frequency appears in a thermal factor, its effect is minimal compared to the dominant real frequencies. Thus, for our purpose of calculating the attempt frequency, this small imaginary mode does not impact the conclusions.

Based on Eq.~\ref{eq:tau}, the calculated attempt frequency $\tau_0$ is found to be $6.12 \times 10^{12}$~Hz. Using this value, the diffusion coefficient $D$ and transport mobility $\mu$ are determined to be $D = 3.59 \times 10^{-7}$~cm$^2$/s and $\mu = 1.37 \times 10^{-5}$~cm$^2$/V$\cdot$s. Despite the strong electronic coupling $V_\mathrm{AB}$, the calculated hole polaron mobility is relatively low at room temperature. This value is similar to that found in nontronite Fe$_2$Si$_4$O$_{10}$(OH)$_2,$\cite{Alexandrov} but lower than values reported in hematite\cite{Adelstein} and rutile TiO$_2$,\cite{Dupuis} which typically range from $10^{-4}$ to $10^{-2}$~cm$^2$/V$\cdot$s. Future experimental studies will be essential to confirm the theoretical prediction in NaMn$_3$O$_7$.

The relatively slow transport arises not from weak electronic interaction but rather from the moderate activation barrier of 0.242 \text{eV} obtained from our DFT calculations. According to Marcus theory, the polaron hopping rate depends exponentially on the activation barrier, even under strong coupling conditions. This reflects a situation where, although the charge transfer pathway is electronically facile, a substantial lattice reorganization is still required to complete the hop. Similar behavior has been observed in $\alpha$-Fe$_{2}$O$_{3}$, where adiabatic transfer coexists with activation energies of 0.18--0.22 \text{eV} depending on hopping direction.\cite{Adelstein}

To further evaluate the impact of temperature on transport performance, we computed the temperature-dependent polaron mobility over the 200–500~K range based on the Einstein relation and Marcus theory, under the assumption of a temperature-independent activation energy $\Delta E^{\ddagger}$:

\begin{equation}
\mu(T) = \frac{e a^2 n \tau_0}{2 k_B T} \exp\left( -\frac{\Delta E^{\ddagger}}{k_B T} \right)
\end{equation}

\begin{figure}[h]
\centering
\includegraphics[width=3.5in]{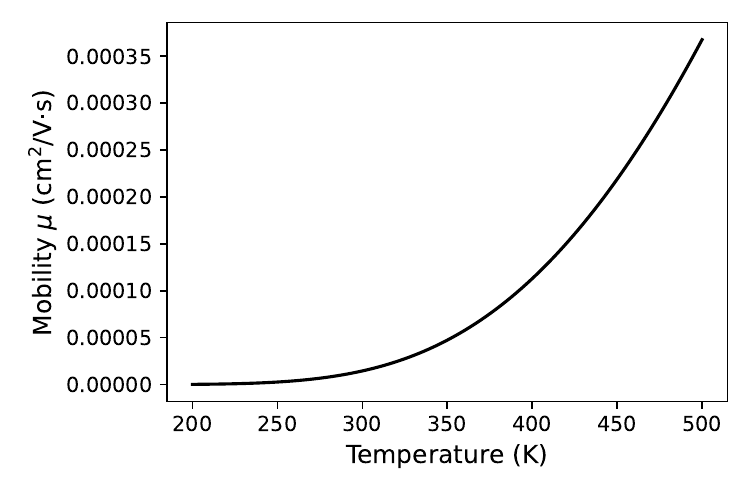}
\caption{Calculated temperature dependence of the oxygen hole polaron mobility $\mu(T)$ in NaMn$_3$O$_7$.}
\label{fig:muT}
\end{figure}

As shown in Fig.~\ref{fig:muT}, the computed polaron mobility $\mu(T)$ increases exponentially with temperature over the range of 200–500~K. At room temperature (298~K), the value is $1.37 \times 10^{-5}$~cm$^2$/V$\cdot$s, and it rises to $3.6 \times 10^{-4}$~cm$^2$/V$\cdot$s at 500~K—an increase of more than one order of magnitude. This result indicates a significant enhancement in polaron transport efficiency at elevated temperatures. 

The exponential increase reflects the Arrhenius-type dependence in the mobility expression, governed by the fixed activation energy barrier of 0.242~eV. Despite the relatively strong electronic coupling in this system, the finite barrier limits low-temperature transport, while high-temperature mobility is significantly enhanced.

\subsection{Oxygen hole polaron transport in Na$_{1.5}$Mn$_3$O$_7$}

After establishing the oxygen hole polaron migration mechanism and its transport mobility in NaMn$_3$O$_7$, we now turn our attention to the oxygen hole polaron transport behavior in the less desodiated phase, Na$_{1.5}$Mn$_3$O$_7$.

\begin{figure}
\centerline{\includegraphics[width=3.6in]{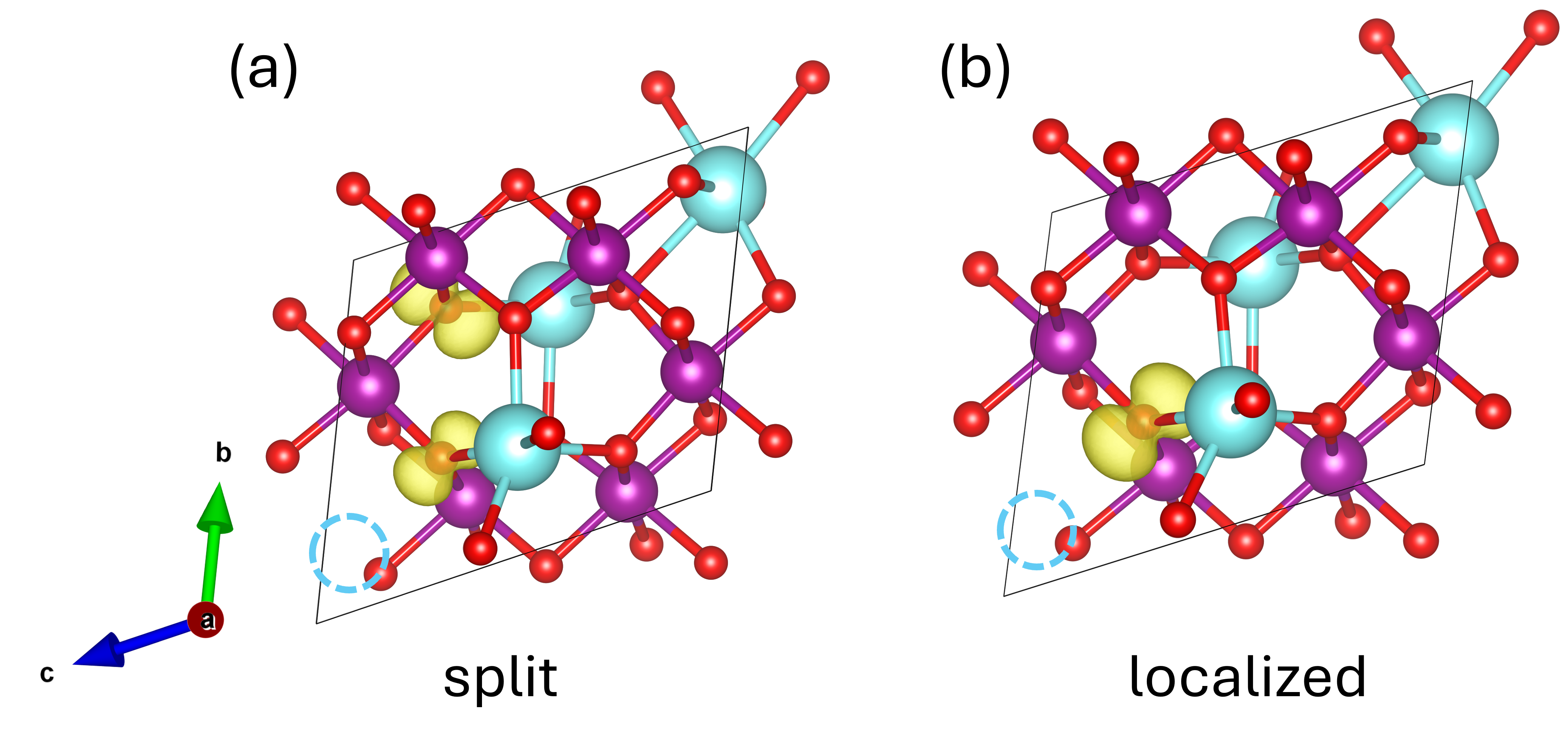}}
\caption{\label{fig:less}(a) Split oxygen hole polaron and (b) localized oxygen hole polaron configuration in Na$_{1.5}$Mn$_3$O$_7$. The Na vacancies are shown in dashed blue circles.}
\end{figure}

In Na$_{1.5}$Mn$_3$O$_7$, there is a single sodium vacancy per unit cell. Despite the reduced number of Na vacancies, DFT calculations reveal that the split polaron configuration still forms, as shown in Fig.\ref{fig:less}. Additionally, a localized polaron configuration can be observed, where the hole polaron is confined to a single oxygen atom near the sodium vacancy.

To compare the relative stabilities of these two configurations, we performed DFT calculations using the Heyd--Scuseria--Ernzerhof (HSE) hybrid functional. The results indicate that the split polaron configuration is approximately 502 \text{m\text{eV}} more stable than the localized polaron configuration.

Based on these findings, we propose that the hole polaron transport mechanism in Na$_{1.5}$Mn$_3$O$_7$ mirrors that observed in NaMn$_3$O$_7$. Specifically, the split polaron configuration serves as both the initial and final state, while the localized polaron configuration corresponds to the transition state. Therefore, the activation energy barrier for hole polaron migration is determined by the energy difference between these two configurations, measured at 502 \text{m\text{eV}}.

\section{Conclusion}

We have demonstrated, through hybrid-functional density functional theory calculations, that oxygen hole polarons in desodiated Na$_{2}$Mn$_{3}$O$_{7}$ undergo long-range migration via a Grotthuss-type mechanism. This process involves alternating transitions between a symmetric split-hole configuration and a localized configuration, resulting in strong electronic coupling characteristic of adiabatic small-polaron hopping.

The split-hole polaron, delocalized over a pair of lattice oxygen atoms without forming an O--O dimer, represents a novel class of polaronic quasiparticles distinct from both traditional small polarons and covalent peroxide species. In contrast to known oxygen polaron migration mechanisms, which typically involve localized hopping, this mechanism proceeds through a dynamically reconfigurable oxygen network enabled by the ordered vacancy structure of the host lattice.

The stabilization of this exotic transport behavior is attributed to the unique electrostatic environment created by the Mn vacancy network and the topological openness of the layered structure. The mechanism persists across a range of Na compositions, indicating its robustness and generality. While our study highlights the role of ordered cation vacancies in enabling split-hole polaron formation and migration, we note that materials exhibiting negative charge-transfer character can also stabilize oxygen-centered holes and may support multi-center delocalization.\cite{Bisogni2016,Seo2016} Moreover, cathodes with ribbon-like ordering patterns, which exhibit strong charge–defect interactions, could also stabilize multi-center delocalization.\cite{house2023delocalized, yu2024ribbon} Additionally, point defects arising from chemical doping can induce charge delocalization phenomena, as observed in Al-doped BaTiO$_3$.\cite{Possenriede} Although these possibilities remain to be systematically explored, they suggest broader principles for identifying materials that may host polaronic species beyond the specific case of ordered-vacancy systems.

Taken together, these results establish a new microscopic picture of oxygen polaron dynamics that extends beyond conventional models and provide guiding principles for the design of redox-active oxides where oxygen hole transport is desirable. In particular, systems featuring undercoordinated oxygen environments and ordered cation vacancies may serve as promising hosts for split-polaron-based transport. Our findings offer theoretical insights relevant to ionic-electronic coupling, oxygen redox reversibility, and functional oxide design in batteries, photocatalysis, and correlated electron systems.

\begin{acknowledgments}
The authors would like to thank Das Pemmaraju for the insightful discussion. This work was supported by the MIT Startup Fund. The computational work used Expanse at San Diego Supercomputer Center through allocation MAT230005 from the Advanced Cyberinfrastructure Coordination Ecosystem: Services \& Support (ACCESS) program, which is supported by National Science Foundation grants \#2138259, \#2138286, \#2138307, \#2137603, and \#2138296.
\end{acknowledgments}

\bibliography{main}
\end{document}